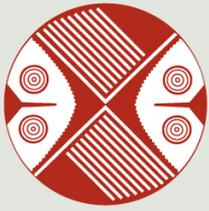
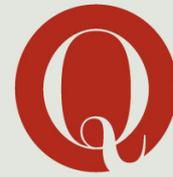

Rodríguez, Facundo

# La materia oscura en el contexto de la teoría del actor-red

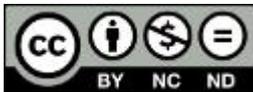

# LA MATERIA OSCURA EN EL CONTEXTO DE LA TEORÍA DEL ACTOR-RED*

*Facundo Rodríguez***

> No sirve de nada sostener que los científicos que se dedican a las ciencias naturales también se la pasan agregando entidades ocultas para encontrar sentido a los fenómenos.
> Bruno Latour

**RESUMEN**

El objetivo de este trabajo es utilizar algunas categorías de la teoría del actor-red desarrolladas por Bruno Latour en el libro *Reensamblar lo social: una introducción a la teoría del actor-red*, para analizar una controversia actual que se presenta en el área de la cosmología: la existencia de la materia oscura y sus características. La idea central que atraviesa este artículo es intentar descifrar qué tipo de agente es la materia oscura y abrir la discusión al respecto desde una perspectiva sociológica.

palabras clave: teoría del actor red – científicos – astronomía – materia oscura

## INTRODUCCIÓN

El presente trabajo busca establecer la pertinencia de considerar a la materia oscura como un actante. Para cumplir con este objetivo realizaremos, en primer lugar, una breve caracterización de la compleja red de actores y actantes de la que forma parte, y, posteriormente, presentaremos y analizaremos la controversia. Es decir que partiremos de la astronomía para describir los desafíos que la materia oscura impone a los estudios cosmológicos e intentaremos aplicar algunas nociones de la teoría del actor-red (TA-R). Para esto último, nos basaremos en el texto *Reensamblar lo social: una introducción a la teoría del actor-red*, de Bruno Latour, en el cual el autor realiza un cuidadoso desarrollo sobre cómo rastrear las asociaciones en el marco de la TA-R.

Podríamos empezar por preguntarnos por qué estudiar esta controversia desde la sociología de las ciencias. En el marco de la TA-R, la respuesta es muy sencilla: las controversias proveen de un recurso esencial para poder establecer cuáles son las conexiones sociales. Además, según Latour, "el relevamiento de controversias científicas respecto de cuestiones de interés debería permitirnos renovar de arriba abajo la escena misma del empirismo y, por lo tanto, la divisoria entre 'natural' y 'social'" (2008: 167).

Seguidamente, podríamos cuestionar el carácter social de un análisis acerca de algo que es no humano y, en este caso, la respuesta puede expresarse nuevamente en palabras de Latour: "Si aceptamos aprender también de las controversias acerca de los no humanos, pronto advertimos que las cuestiones de hecho no describen qué tipo de agencias pueblan el mundo mejor de lo que las palabras 'social', 'simbólico' y 'discursivo' describen qué es un actor humano y los extraños que lo dominan" (2008: 161). Además, agrega:

> la TA-R no es la afirmación vacía de que son los objetos los que hacen las cosas "en lugar de" los actores humanos: dice simplemente que ninguna ciencia de lo social puede iniciarse siquiera si no se explora primero la cuestión de quién y qué participa en la acción, aunque signifique permitir que se incorporen elementos que, a falta de mejor término, podríamos llamar no-humanos (Latour, 2008: 107).

Lo que intentaremos hacer a continuación es describir la problemática planteada realizando una breve reseña de cómo surgió y por qué aún se encuentra vigente la discusión acerca de la existencia de la materia oscura. Luego, intentaremos analizar dicho relato para establecer los actores intervinientes, los actantes, la agencia de cada uno y los grupos que se conforman en torno a esta controversia.

**RELATO: LA MASA INVISIBLE**

Cuando los científicos estudian los astros, la mayor parte de la información que pueden recabar está en la luz que estos emiten o absorben, es decir, de los datos que brinda la radiación proveniente de los planetas, estrellas, galaxias, etc., y también aquella que proviene de etapas tempranas del universo. Las imágenes que se toman en astronomía son mediciones de la cantidad y el tipo de luz que emite aquello que observan. Por ejemplo, se puede determinar la cantidad de luz visible que emite una galaxia, de ondas de radio o de rayos X, según lo que se quiera estudiar. La radiación, a través de sus características como intensidad, longitud de onda, polarización, líneas espectrales, entre otras, aporta datos indirectos de los objetos estudiados tales como la masa o la temperatura. Esta información es abordada de diferente forma por los investigadores y, según lo que se estudie y de la manera en que se lo haga, surgen diferentes grupos. Por ejemplo, los que estudian la dinámica de los planetas, la evolución de las estrellas, la estructura de nuestra galaxia, los procesos de altas energías, los radio-astrónomos, los que hacen simulaciones de formación de galaxias o los que estudian la estructura en gran escala del universo, por mencionar algunas categorías posibles en las que pueden agruparse dichos científicos.

Debido a la discrepancia entre datos obtenidos mediante la observación y los modelos físicos, surgió la idea de que puede haber un tipo de materia muy diferente de la que forman los planetas y las estrellas. Al parecer, hay una gran cantidad de materia que está presente en las grandes estructuras del universo y que no emite radiación pero, como toda masa, ejerce gravedad y genera atracción y efectos en el espacio. Esto es lo que los físicos y astrónomos denominan "materia oscura", y estudian para intentar explicar si realmente existe, qué tipos de partículas la componen y por qué no emite radiación.

Como acabamos de mencionar, este es un tema que aún no está cerrado y que genera grandes controversias. Por lo tanto, atendiendo la limitación de que todavía pueden aparecer nuevos actores, en este trabajo intentaremos hacer una aproximación al análisis de esta controversia a partir de la TA-R.

En primer lugar, es necesario destacar que la materia oscura solo se evidencia a gran escala, es decir, en las investigaciones que se realizan sobre el sistema solar, o, individualmente, las estrellas que componen nuestra galaxia, no es necesario que los investigadores incorporen a la materia oscura. Pero, en la década de 1920, cuando se hicieron los primeros trabajos que intentaban estimar la cantidad total de materia en el universo, se encontró que la materia observada no era suficiente para explicar los efectos gravitatorios.[1] Tanto las estrellas de nuestra galaxia que giran alrededor del centro galáctico como las galaxias que giran alrededor de un cúmulo parecen estar contenidas en una estructura invisible de mayor tamaño. A continuación, vamos a tomar algunos ejemplos y analizaremos la necesidad de la materia oscura para explicarlos.

Uno de ellos, posiblemente el más claro, es el movimiento de las galaxias espirales. Este tipo de galaxias se caracteriza porque las estrellas y el polvo que la componen se distribuyen principalmente sobre un plano y presentan un centro muy denso del cual emergen brazos espirales. En la figura 1, vemos una galaxia espiral NGC 1232. Este tipo de galaxias, debido a su importante movimiento de rotación, fue el primero que permitió determinar la velocidad de rotación de las estrellas que la componen.

**Figura 1. Galaxia espiral llamada NGC 1232**

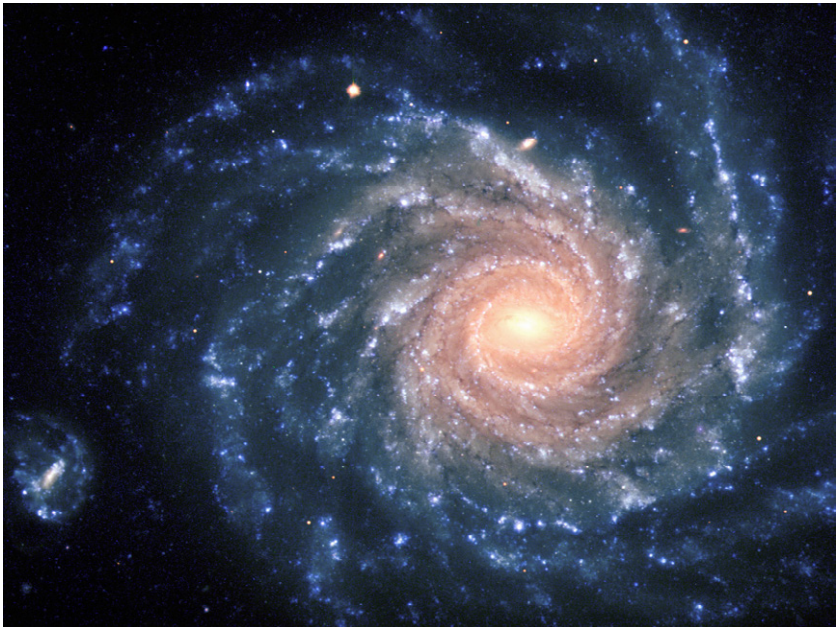

*Fuente*: Observatorio Europeo Austral.

[1] James Jeans y Jacobus Kapteyn indicaron en 1922 que debía haber aproximadamente tres estrellas no detectables por cada una de las registradas. En 1932, esto fue confirmado por Jan Henrik Oort. Sin embargo, el primero en pensar en la materia oscura fue Fritz Zwicky, en 1933, quien además acuñó ese término.

Las galaxias rotan alrededor de su centro y esto se traduce en la luz que recibimos de ellas. Más precisamente, en la variación de la longitud de onda; podría decirse que una mitad de la galaxia se ve un poco más roja y la otra mitad un poco más azul, porque la rotación hace que parte de la radiación proveniente de la galaxia incremente su longitud de onda y otra sufra un decrecimiento –debido al movimiento relativo entre la estrella que emite esa radiación y el observador–. Esto es conocido como *efecto Doppler*. Las diferencias de colores permiten determinar la velocidad que posee cada punto.

En la década de 1970, los astrónomos pudieron comenzar a estimar las velocidades de rotación en función de la distancia al centro de la galaxia, y se esperaba que, si había una gran concentración de materia en el centro y los brazos eran menos densos, la velocidad medida debía disminuir al aumentar la distancia al centro; pero, sorprendentemente, como observamos en la figura 2, esto no fue así: ¡las velocidades se mantienen constantes![2]

Esto dio origen a una importante controversia debido a las diferentes interpretaciones posibles, que iban desde fallos en las observaciones a cuestionamientos a la universalidad de la relatividad general, pasando por la hipótesis de que podría haber una gran cantidad de materia que no vemos y que este disco podría estar contenido en una especie de nube de mucho mayor tamaño y masa pero que no emitiría radiación. Esta última propuesta fue la que prevaleció, y a este componente del universo se lo denominó *materia oscura*, ya que hace referencia a que no emite luz.

Tanto los físicos teóricos como los astrónomos observacionales supusieron que, si había una gran cantidad de masa invisible, en el sentido de que no emitía radiación, esto debía generar otros efectos perceptibles que brindaran un poco más de información. Así, por ejemplo, cuando estudiaron grandes cúmulos de galaxias, es decir, cientos de galaxias que giran alrededor de un mismo centro de masa, observaron que estos cúmulos están rodeados de un halo, una especie de nube difusa, de gas muy caliente que emite en rayos X. La masa de ese gas no era suficiente para explicar la dinámica de estos cúmulos, pero la distribución que este gas posee y su tempe-ratura parecen afirmar los indicios sobre la presencia de una gran cantidad de materia oscura. Estas observaciones sugirieron que el cúmulo en su conjunto está embebido en un gran halo de materia oscura que retiene al gas.[3]

---

[2] En realidad, desde hacía muchos años se podían medir las curvas de rotación de las galaxias. Horace Babcock, por ejemplo, ya en 1939 había estimado la curva de rotación de la galaxia de Andrómeda (M31), pero durante décadas los resultados eran contradictorios. Debido a las mejoras en las técnicas de medición, estas determinaciones fueron cada vez más precisas y, en 1975, Vera Rubin y Kent Ford presentaron resultados contundentes al obtener curvas de rotación planas con un grado de precisión muy superior.

**Figura 2. Velocidad de rotación de las estrellas en función de su distancia al centro de la galaxia**

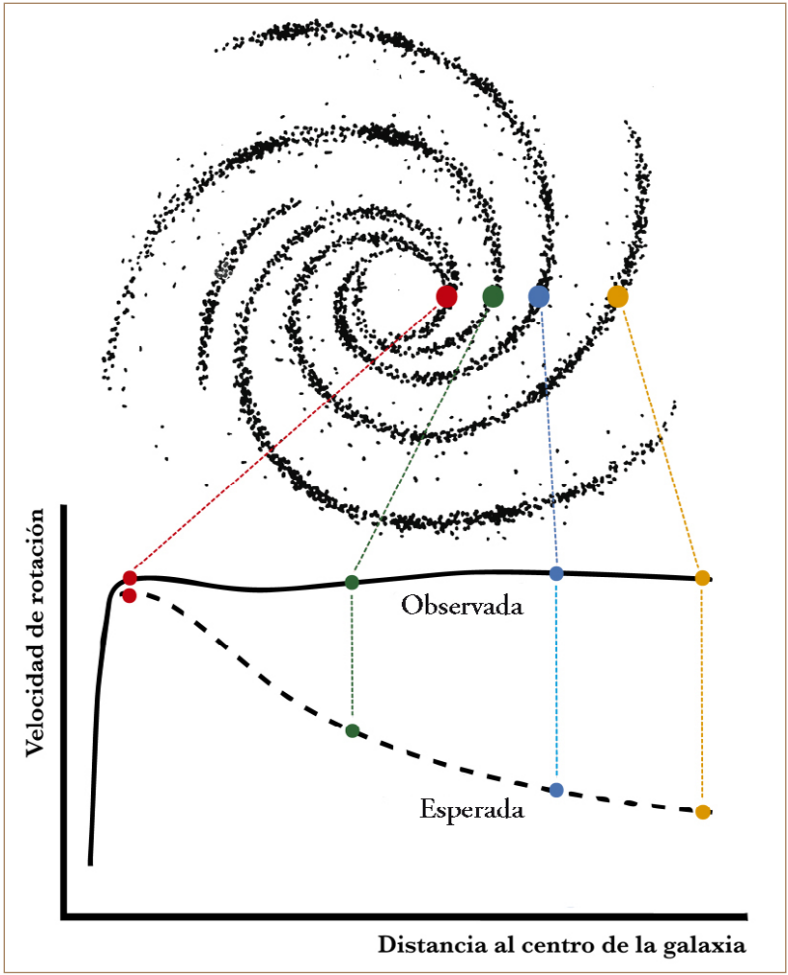

*Fuente*: Sureda (2015).

[3] El proyecto Chandra, del que forma parte el telescopio espacial de rayos X, por ejemplo, tiene como uno de sus objetivos estudiar la relación entre los rayos X y la materia oscura.

Como se obtienen pruebas indirectas de su existencia, pero aún no se han podido determinar qué tipo de partículas pueden componer esta extraña clase de materia, se proponen nuevas observaciones para poder establecer si, por ejemplo, hay que modificar la teoría de la gravedad, como afirmaron algunos científicos desde el surgimiento de esta controversia –por esto se siguen buscando modelos teóricos alternativos que no hagan uso de la materia oscura–. Otros científicos se preguntan si realmente existe, dado el hecho de que aún no se ha podido generar consenso sobre ninguna de las propuestas sobre el tipo de partículas que la forman.[4]

Una de las pruebas más interesantes que ha superado es la de las lentes gravitacionales. Desde que se estableció que los cúmulos estaban embebidos en una gran masa "transparente" y cuya radiación no es perceptible, los científicos –y en especial los físicos relativistas– postularon que esto debería generar una curvatura en el espacio-tiempo de manera tal que la imagen de algún objeto que se encuentra detrás del cúmulo se vería afectada; es decir, la materia oscura actuaría como una especie de lente.[5] Muchos astrónomos consideraban esta idea como descabellada, ya que les parecía imposible que hubiera evidencia observacional de esto, pero la materia oscura superó esta prueba y, tal como vemos en la figura 3, es lo que sucede. Es más, los objetos más lejanos que se pueden estudiar hoy en día son fruto del uso de los cúmulos como una especie de telescopios naturales.

Obviamente, esta gran cantidad de masa que no vemos tiene importantes implicancias en la manera de entender en el ámbito de la cosmología.[6]

---

[4] Un ejemplo de esto podría ser la búsqueda de objetos compactos que emiten baja radiación, tales como agujeros negros, enanas marrones o rojas y estrellas de neutrones. Algunos de los proyectos dedicados a la búsqueda de eventos de este tipo son MACHO, EROS y OGLE, pero sus resultados indican que las cantidades de materia encontrada debido a estos objetos es insuficiente para reemplazar a la materia oscura, debido a que es muy bajo su aporte al total de materia.

[5] Einstein, como parte del planteo de la relatividad general, ya había pensado que la curvatura del espacio-tiempo podía generar efectos similares a la lentes y curvar la luz. En 1919, Arthur Eddington determinó observacionalmente la curvatura que se producía en la trayectoria de la luz proveniente de estrellas distantes al pasar cerca del Sol, y demostró la existencia de este fenómeno. En 1937, Fritz Zwicky publicó un artículo en el cual propuso el uso de esto para determinar masas de galaxias o cúmulos de galaxias y poner a prueba la existencia de materia oscura, abriendo una nueva línea de estudios que perdura con fuerza en la actualidad, combinando estudios teóricos con observaciones.

[6] Da cuenta de esto el modelo cosmológico más aceptado actualmente, denominado ΛCDM, y que toma como uno de sus parámetros fundamentales la cantidad de materia oscura. Además de las determinaciones de los parámetros cosmológicos realizados, por ejemplo, por los proyectos WMAP y Planck, que corroboran este modelo.

Y si bien en estas escalas se trabaja con la radiación proveniente de cuerpos lejanos, se pueden comparar las observaciones obtenidas con los modelos teóricos mediante la utilización de simulaciones numéricas realizadas con potentes sistemas de cómputos.[7] Al parecer, estos presentan un muy buen acuerdo y la materia oscura sería uno de los componentes esenciales del universo, que posibilitó la formación de las grandes estructuras observadas actualmente. Es decir, de nuevo, la materia oscura superó la prueba de ser necesaria para reproducir las estructuras actuales –usando los modelos cosmológicos más aceptados–, ya que, si los científicos que realizan las simulaciones no la tienen en cuenta, no logran formar los grandes aglomerados de galaxias observados.

**Figura 3. Imagen de lentes gravitacionales tomadas por el Telescopio Espacial Hubble**

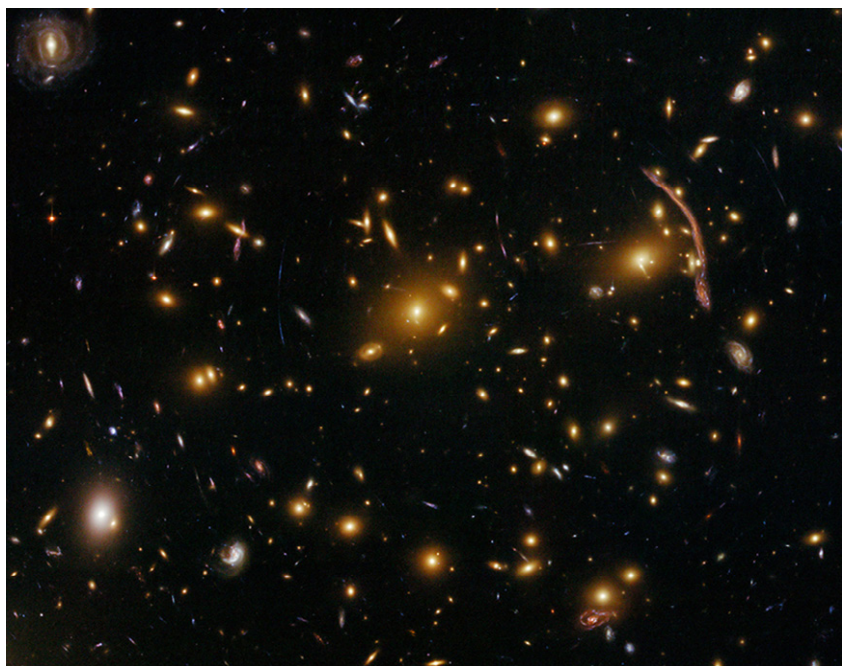

*Fuente*: NASA, ESA, S. Beckwith (STScI) y el equipo HUDF. Disponible públicamente en el repositorio de imágenes del telescopio espacial Hubble: <http://hubble.org>.

[7] Diversos grupos, mediante la comparación entre simulaciones como Millennium, realizada por la sociedad Max Planck, y catálogos observacionales de galaxias, como el relevamiento del cielo llevado a cabo por el proyecto Sloan, intentan relacionar las propiedades de galaxias con las de la materia oscura.

Hasta ahora parece que, a pesar de que no se sabe qué es, las pruebas de su existencia son sólidas. Claro que esto es solo una versión de los hechos, pero no la única: muchos astrónomos y físicos que estudian la cosmología se centran en el estudio del espacio-tiempo y creen que, más allá de los resultados obtenidos hasta el presente, la materia oscura es la prueba de que la teoría de la relatividad general debe ser modificada o reemplazada y proponen modelos teóricos para establecer qué parte de la teoría sería la que está "fallando".[8]

Otros científicos intentan encontrar partículas que expliquen esas propiedades de la materia oscura, pero a pesar de haber realizado varias propuestas, hasta el momento ninguna ha logrado un consenso general.[9]

### ANÁLISIS: EL ACTANTE INVISIBLE

A partir del relato anterior, se pueden identificar los actores o actantes que participan de la controversia. Latour señala una forma clara para poder estudiar esto:

> Si nos mantenemos en nuestra decisión de partir de nuestras controversias sobre actores y agencias, entonces cualquier cosa que modifica con su incidencia un estado de cosas es un actor o, si no tiene figuración aún, un actante. Por lo tanto, las preguntas que deben plantearse sobre cualquier agente son simplemente las siguientes: ¿Incide de algún modo en el curso de la acción de otro agente o no? ¿Hay alguna prueba que permita que alguien detecte esta incidencia? (Latour, 2008: 106).

De la descripción anterior, se puede identificar fácilmente la presencia de varios agentes, entre ellos los científicos –en conjunto, como individuos y como parte de los grupos o líneas de investigación de los que forman parte–, la radiación electromagnética, los telescopios y otros instrumentos de observación, los sistemas de cómputo, las estrellas, el gas, las galaxias, los cúmulos de galaxias, la teoría de la relatividad, los diversos modelos teóricos y la materia oscura.

---

[8] A lo largo de todo el mundo, hay grupos que estudian actualmente modelos alternativos a la relatividad general, tales como la teoría de la relatividad modificada de tensor-vector-escalar, la teoría gravitacional no-simétrica, la teoría de expansión cósmica en escala, entre otras.

[9] Diversos grupos se encuentran estudiando partículas tales como las WIMP –denominadas así por sus siglas en inglés, estas tienen masa pero interactúan débilmente–, axiones, neutrinos estériles, etcétera.

La exposición de las acciones, en el contexto de la TA-R, es un poco más compleja; por lo tanto, lo que intentaremos hacer es simplemente esbozar un análisis de estas acciones, ya que, según Latour:

> la acción no se realiza bajo el pleno control de la conciencia; la acción debe considerarse en cambio como un nodo, un nudo y un conglomerado de muchos conjuntos sorprendentes de agencias y que tienen que ser desenmarañados lentamente. Es esta venerable fuente de incertidumbre a la que queremos dar vida nuevamente con la extraña expresión actor-red (Latour, 2008: 70).

Por otro lado, valiéndonos de otras palabras de este mismo autor: "Si se menciona una agencia, hay que presentar el relato de su acción, y para hacerlo hay que explicitar más o menos qué pruebas han producido qué rastros observables" (2008: 82). Entonces, podemos describir a grandes rasgos las acciones realizadas por algunos agentes.

Los astrónomos y físicos realizan una gran variedad de acciones relacionadas con la investigación del universo –observar, medir, estimar, etc.– y todas tienen correlatos observables, comprobación y generación de teorías, discusiones académicas respecto de estas, conformación de grupos, etc. Entre tales acciones se encuentra la de "descubrir" la materia oscura y ponerla a prueba para evidenciar su existencia o no. De esta manera, podemos afirmar que los científicos que trabajan en áreas relacionadas con astronomía y cosmología son actores de esta controversia. Pero no solo lo son como conjunto, sino que también son relevantes las diferentes tareas que realizan, ya que algunos se dedican a hacer observaciones o detecciones indirectas de la materia oscura, mientras que otros –por ejemplo, los que trabajan en escalas planetarias o estelares– pueden realizar todo su trabajo sin tomar en cuenta su presencia. También hay quienes realizan simulaciones, formulaciones o evaluaciones de modelos teóricos, etc. Las diferentes tareas hacen que conformen una red de grupos de investigadores que, a su vez, se relacionan con los demás agentes a partir del rol que cumplen dentro de la división del trabajo científico. Consideramos importante visualizar a los científicos, desde el comienzo, como un grupo heterogéneo con diferentes perspectivas, posiciones y relaciones respecto de la materia oscura.

A continuación, mencionaremos acciones realizadas por algunos agentes con la finalidad de bosquejar roles dentro de la red.

Las estrellas emiten radiación, que luego es detectada y estudiada, giran en torno al centro galáctico, conforman las galaxias. Las galaxias residen –y se forman– en el interior de los halos de materia oscura, rotan, conforman cúmulos de galaxias. El gas se expande alrededor de las galaxias y cúmulos de galaxias, es retenido por la materia oscura, se calienta y emite radiación, que luego es analizada. Los telescopios y otros instrumentos de observación permiten la detección de la radiación proveniente de las estrellas, galaxias, cúmulos, gas, etc., y también producen datos que permiten relacionar las teorías con los astros observados y la materia oscura. Por ejemplo, obtener imágenes que verifiquen especulaciones teóricas como en el caso de las lentes gravitacionales. Los sistemas de cómputo permiten realizar simulaciones y evaluar si sus resultados se corresponden con la presencia de materia oscura.

Los modelos teóricos plantean nuevos retos a la materia oscura e intentan describir la realidad observada.

La materia oscura, por su lado, también "acciona". Una de las acciones que tomamos como más relevantes es la de evidenciarse solo a escalas suficientemente grandes. Otras son las de no emitir radiación, interactuar gravitacionalmente con las galaxias, grupos y cúmulos de galaxias, superar las pruebas que los científicos imponen y, así, hacer que estos conformen grupos.

Entonces, se puede comenzar a confeccionar una red con actores –los científicos que pertenecen a los diferentes grupos antes mencionados– y actantes: estrellas, galaxias, cúmulos de galaxias, gas, instrumentos, sistemas de cómputos, modelos teóricos y la materia oscura –podrían ser más, pero son los que tomamos para este breve análisis. Ateniéndose a la descripción que hemos realizado de esta controversia, se puede plantear que dentro de esta red tienen un papel preponderante los científicos, como actores, y la materia oscura, como actante.

Dentro del grupo de los científicos, están aquellos que intentan demostrar que la materia oscura existe y la utilizan tanto para los cálculos como para las simulaciones y, por otro lado, aquellos que intentan demostrar que esta aparece debido a la falta de entendimiento de la relatividad general o de la necesidad de nuevas teorías que expliquen la dinámica del universo a gran escala. Esto nos permite retomar nuevamente las palabras de Latour, quien afirma que "los actores también se dedican a criticar a otras agencias acusándolas de ser falsas, arcaicas, absurdas, irracionales, artificiales o ilusorias" (2008: 87). En este caso, sería el grupo de científicos que descreen de la materia oscura el que intenta decir que este agente es ilusorio o artificial. Y es allí donde toma un rol activo la materia oscura, ya que, a pesar de que muchos descreen de su existencia, supera las pruebas que le imponen, aún no puede ser reemplazada por otro actante o desaparecer, dejando de cumplir un rol activo porque ya no es necesaria en alguna nueva teoría.

Según Latour, "así como los actores constantemente son llevados a participar en la formación y destrucción de grupos [...], también se dedican a proveer explicaciones polémicas de sus acciones así como de las de los demás" (2008: 75). En este caso, los actores que están en contra y a favor de la materia oscura forman grupos y, a través de artículos científicos, proveen explicaciones sobre las posiciones que toman, e intentan darle más o menos jerarquía a la materia oscura. A su vez, podemos decir que es la materia oscura, como actante, la que genera la conformación de estos grupos, al estar constantemente afirmándose como entidad, al superar las pruebas impuestas pero sin revelar totalmente su composición y, por lo tanto, seguir activamente generando controversia.

## CIERRE

La polémica aún está abierta y los científicos van tomando nuevas posturas, conformando nuevos grupos y generando nuevos discursos y prácticas. Por lo tanto, es difícil establecer cuál de todas las posiciones planteadas será las más aceptada, luego de que pase algún tiempo y se sumen estudios, discusiones y consensos.

De todas maneras, la principal idea que este trabajo buscó presentar es que se puede pensar la materia oscura como un actante y que, según cuál sea el curso de las investigaciones, puede establecerse como tal, tomar más importancia o, por el contrario, ser reemplazada por otro agente o simplemente desaparecer al dejar de tener un rol activo en la ciencia.

## BIBLIOGRAFÍA